\documentclass[letterpaper,10pt]{sig-alternate-10pt}

\usepackage{amssymb}
\usepackage{graphicx}
\usepackage{url}

\newcommand{\cc}{\ensuremath{\mbox{cc}}}
\newcommand{\tr}{\ensuremath{\mbox{tr}}}
\newcommand{\ie}{{\em i.e.}}
\newcommand{\traceroute}{{\tt traceroute}}

\newcommand{\myitem}{\noindent$\bullet$\ }

\newcommand{\inet}{{\em \sc inet}}
\newcommand{\web}{{\em \sc web}}
\newcommand{\ptp}{{\em \sc p2p}}
\newcommand{\ip}{{\em \sc ip}}

\renewcommand{\subsubsection}[1]{\medskip\noindent{\em #1}\smallskip}

\newcommand{\plots}[3][0.35]{\begin{figure}[!h]
\includegraphics[scale=#1]{Plots/Inet/#2.fig.eps}
\hfill
\includegraphics[scale=#1]{Plots/P2P/#2.fig.eps}\\
\includegraphics[scale=#1]{Plots/Web/#2.fig.eps}
\hfill
\includegraphics[scale=#1]{Plots/IP/#2.fig.eps}
\caption{#3
From left to right and top to bottom: \inet, \ptp, \web\ 
and \ip\ graphs.}
\label{fig_#2}
\end{figure}}

\begin{document}

\title{Measuring Fundamental Properties of\\ Real-World Complex Networks}
\author{
\alignauthor
Matthieu Latapy \hspace*{0.1cm} and \hspace*{0.1cm} Clemence Magnien\\
\affaddr{LIP6 -- CNRS and University Pierre \& Marie Curie}\\
\affaddr{4 place Jussieu, 75005 Paris, France}\\
\email{firstname.lastname@lip6.fr}
}

\maketitle

\begin{abstract}

Complex networks, modeled as large graphs, received much
attention during these last years. However, data on such
networks is only available through intricate measurement
procedures. Until recently, most studies assumed that these
procedures eventually lead to samples large enough to be
representative of the whole, at least concerning some key
properties. This has crucial impact on network modeling and
simulation, which rely on these properties.

Recent contributions proved that this approach may be
misleading, but no solution has been proposed. We provide
here the first practical way to distinguish between cases
where it is indeed misleading, and cases where the observed
properties may be trusted. It consists in studying how the
properties of interest evolve when the sample grows, and in
particular whether they reach a steady state or not.

In order to illustrate this method and to demonstrate
its relevance, we apply it to data-sets on complex network
measurements that are representative of the ones commonly used.
The obtained results show that the method fulfills its goals
very well. We moreover identify some properties which
seem easier to evaluate in practice, thus opening interesting
perspectives.

\end{abstract}

\medskip

\section{Context.}
\label{sec_context}

Complex networks of many kinds, modeled as large graphs, appear in
various contexts. In computer science, let us cite internet maps (at
IP, router or AS levels, see for instance
\cite{faloutsos99sigcomm,gkantsidis,magoni01analysis,skitterurl}),
web graphs
(hyperlinks between pages, see for instance \cite{kleinberg99web,broder00graph,boldi04www,boldi04dcc,webgraphurl}),
or data exchanges (in peer-to-peer systems, using e-mail, etc, see for
instance \cite{lefessant04exploiting,voulgaris04exploiting,leblond2004p2p,iptps05}).
One may also cite many examples among social, biological or linguistic
networks, like co-authoring networks, protein interactions, or
co-occurrence graphs for instance.

It appeared recently (at the end of the 90s
\cite{watts1998smallworld,faloutsos99sigcomm,kleinberg99web,albert99nature,broder00graph})
that most real-world complex networks have nontrivial properties
which make them very different from the models used until then
(mainly random, regular, or complete graphs and ad hoc
models). This lead to the definition of a set of statistics,
the values of which are considered as fundamental properties of the complex
network under concern. This induced in turn a stream of studies aimed at
identifying more such properties, their
causes and consequences, and capturing them into relevant models.
They are now used as key parameters in the study of
various phenomena of interest like robustness \cite{albert00error,kim2004random},
spreading of information or viruses \cite{pastor01epidemic,ganesh2005epidemics}, and 
protocol performance \cite{li2004overlay,lefessant04exploiting,voulgaris04exploiting,iptps05}
for instance. They are also the basic parameters of many network
models and simulation systems, like for instance {\sc brite} \cite{medina01brite}.
This makes the notion of fundamental properties of complex networks
a key issue for current research in this field.
For recent surveys on typical properties and related issues, see for instance
\cite{brandes05lncs,bornholdt03book}.

However, most real-world complex networks
are not directly available: collecting data about them requires the use of
a measurement procedure. In most cases, this procedure is an intricate
operation that gives a {\em partial} and possibly {\em biased} view.
Most contributions in the field then rely on the following
(often implicit) assumption:
during the measurement procedure, there is an initial phase in which the
collected data may not be representative of the whole, but {\em when the
sample grows one reaches a steady state where the fundamental properties
do not vary
anymore.} Authors therefore grab a large amount of data (limited by the
cost of the measurement procedure, and by the ability to manage the
obtained data) and then suppose that the obtained view is representative
of the whole, at least concerning these properties.

Until recently, very little was known on the relevance of this
approach, which remains widely used (because in most case there
is no usable alternative method).
This has long been ignored, until the publication of some
pioneering contributions \cite{lakhina02sampling,barford01imc} showing that
the bias induced by measurement procedures is
significant, at least in some important cases. It is now
a research topic in itself, with both theoretical, empirical and experimental
studies; see for instance \cite{lakhina02sampling,barford01imc,achlioptas05bias,guillaume2005metro,dallasta04statistical}\,\footnote{
Note however that, because of its importance
and because its measurement can be quite easily modeled, the case of
internet measurements with \traceroute\ received most attention.}.
In this stream of studies, the authors mainly try to identify the impact
of the measurement procedure on the obtained view and to evaluate the
induced bias. The central idea, first introduced in \cite{lakhina02sampling,barford01imc}, is to
take a graph $G$ (generally obtained from a model or a real-world
measurement), simulate a measurement of $G$ thus obtaining the
view $G'$ and compare $G$ and $G'$. This gave rise to significant
insight on complex network metrology, but much remains to be done.

\section{Approach and scope.}
\label{sec_approach}

Our contribution belongs to the current stream of studies on real-world
complex networks, and more precisely on the measurement of these networks.
It addresses the issue of the estimation of their basic properties,
with the aim of providing a {\em practical} solution to this issue.
Indeed, until now, authors studying real-world complex networks had no choice
but to follow the classical assumption that their sample is large
enough to be representative of the whole, even though this has been
proved to be far from obvious
\cite{lakhina02sampling,barford01imc,achlioptas05bias,guillaume2005metro,dallasta04statistical}.
We will make it possible to evaluate the relevance of this
classical assumption in practical cases.

We notice that the vast majority of real-world complex network
studies rely on samples obtained through a measurement procedure
that is {\em interrupted when the obtained sample
is considered large enough} to be representative of the whole.
Then, we mimic this by processing very large measurements of
real-world complex networks: we study what the 
observed properties would be if one had stopped the measurement when the
sample had reached a given size, smaller than the final one.

The main strength of this approach is that it relies on
{\em real} measurements of complex networks, while previous works
had to model the complex network under concern, the measurement
process, or both, see for instance \cite{lakhina02sampling,achlioptas05bias,guillaume2005metro,dallasta04statistical}. Such a modeling is
a challenging task since the measurement procedure generally is
intricate, and since we do not know the underlying complex
network that we actually measure. We avoid these problems here
since we rely on real-world data, obtained in a way that is
representative of what is done in practice.

This also means that measuring the same complex
networks but in another way, and/or measuring other
complex networks, may lead to different results.
This is why we
paid high attention to use measurements that are representative
of the ones commonly used, and come from four very
different contexts (see Section~\ref{sec_data}); this reduces the
risk of results specific to one case. In each of these
contexts, we moreover used several measurements (of different
sizes, conducted at different times, and/or with significantly
different methods); all the results were consistent and we
present here one typical example for each case. Notice also
that we provide the programs we used here, which makes it possible
to conduct the same analysis on any measurement data-set \cite{dataurlanon}.

\medskip

Before turning to the description of our data-sets and entering
in the core of this contribution, let us emphasize a few key
points.

\myitem
Though we use real-world data in our study, we do not seek
results on these particular examples. It makes no doubt that
studying them in depth would also be relevant, and that our
observations raise interesting issues on each particular case, but
this is not our concern here. We only consider them as typical
large-scale measurements which we use to illustrate our approach.

\myitem
Likewise, we will not discuss the measurement procedures themselves,
which may vary and may be improved; the key point is that these
measurements are representative of the ones used in current
research.
In particular, we follow the classical convention consisting
in ignoring
the bias induced by the fact that the complex network under
concern may evolve during the measurement. This is an important
and interesting issue, but it is out of the scope of this paper.

\myitem
It must also be clear that handling such graphs, together with their
evolution, is an algorithmic challenge. It does not only force us to
use important capacities in central memory and in processing power:
algorithms with a time or space cost more than
linear in the number of nodes $n$ and/or links $m$ are almost unusable in
this context\,\footnote{One may use compression techniques
to reduce central memory requirements, see for instance \cite{boldi04www,boldi04dcc},
or streaming algorithms which make central memory storage unnecessary,
see for instance \cite{jowhari05cocoon,palmer2002tool}, but this is out
of the scope of this paper.}.
We will therefore carefully choose the algorithms we use in our
computations, and discuss their complexities all along the paper\,\footnote{
The given complexities will always be the ones in the worst
case, the notation $\Theta(f(n,m))$ meaning that it is bounded
by $f(n,m)$ and that this bound is tight; instead, $O(f(n,m))$
means that the bound may be weak.
In our cases, $m>n$, therefore we will follow the classical
convention assuming that $m$ is in $\Omega(n)$.}.

\section{Method and data-sets.}
\label{sec_data}

To achieve our goal, we need data in the following form:
given a real-world complex network measurement, for each
integer $n$ we need the graph one would obtain if this
measurement had been stopped as soon as $n$ nodes had been
discovered.
We then compute the properties under concern
for each of these graphs, obtaining plots of their value
as a function of the sample size $n$\,\footnote{To save
computation time, we considered only the values of $n$
in $\{\frac{i*N}{100},\ i=1, \cdots, 100\}$ (where $N$
denotes the number of nodes at the end of the full
measurement) in all the paper, which gives plots with
$100$ points.}.

Our data-sets are derived from raw data on how complex networks
are measured, which we describe below.
They come from some of the largest 
and highest quality data-sets currently available, and
span quite well the variety of complex networks usually
considered in computer science.
From this raw data,
we first extracted, for each node and link, the time at
which it was discovered\,\footnote{
Following the classical conventions in complex network
studies, we removed multiple links (by considering only
the {\em first} time each link is discovered),
and we removed loops (by considering that
discovering a loop $(v,v)$ is equivalent to discovering only the
node $v$).}. Then we wrote a program that runs through
this stream of node and link arrivals (ordered by the time
at which they are discovered) until the sample
reaches the prescribed size $n$, and then computes the desired
statistics.

Because these data-sets and the program may be useful for other purpose, and because
they are needed to reproduce our results, we provide them at \cite{dataurlanon}.

We recall that we only use these data-sets as {\em examples}
here; discussing the relevance of such graphs and their
particular properties is out of the scope of this paper.
The key point is that they are representative of what is used in
most studies, and that in most cases they are significantly
larger. It means that most known results on these objects are
actually derived from samples lying somewhere between the beginning
and the end of the measurement in our cases.

\subsubsection{The \inet\ data-set.}

This data-set comes from the {\em Skitter} project at {\sc caida}
\cite{skitterurl}.
Several machines scattered around the world run
\traceroute-like probes to a list of almost
$1\,000\,000$ destinations, on an approximately daily basis.
They record each route
discovered this way, together with the time at which the probe
was launched (and additional information that we do not need here).
They make this data freely available for academic research.

Such measurements are often used to construct maps of the internet
at IP, router or AS levels. The IP maps are nothing but the set
of all IP addresses viewed during the measurement, with a link
between any two of them if they are neighbors on a collected path.
Obtaining router or AS maps from such data is a challenge in
itself, and subject to some error, see for instance \cite{magoni01analysis}.
Here we will simply consider the IP level.

We downloaded all the data collected by {\em Skitter} from january 2005
to may 2006. During this
period, $20$ machines ran probes with no interruption (other experienced
interruptions, thus we did not include them), leading to
$4\,616\,234\,615$
traceroute-like
records, and approximately $350$ gigabytes of compressed data.
We assumed that the links corresponding to a given route were
seen at the time (in seconds) the probing of this route was started.

The graph finally obtained contains $1\,719\,037$ nodes and
$11\,095\,298$ links.

\subsubsection{The \web\ data-set.}

Web graphs, \ie\ sets of web pages identified by their URL
and hyper-links between them, are often used as typical examples
of complex networks. Indeed, it is quite easy to get large
such graphs using a {\em crawl}: from a set of initial pages
(possibly just one), one follows its links and iterates this
in a breadth-first manner. Collecting huge such graphs however is
much more difficult, since several reasons like limitations
in computing capabilities and crawling policies lead to many
technical constraints.

Here we used a data-set provided by one of the current leading projects on
web crawling and management, nam\-ely {\em WebGraph} \cite{boldi04www,boldi04dcc,webgraphurl}. Their
crawler is one of the most efficient currently running, and
their data-sets on web graphs are the largest available ones.
They provided us with a web graph of pages in the {\tt .uk}
domain containing
$39\,459\,925$
nodes (web pages) and
$921\,345\,078$
directed links (not including loops). Moreover, they provided us with the time
at which each page was visited (each was visited only once),
thus at which each node and its outgoing links were discovered.
This crawl has been ran from the 11-th of
July, 2005, at 00:51, to the 30-th at 23:24, leading to almost
20 days of measurement. The time precision is $1$ minute.

From this data, we obtained a final graph with 
$39\,459\,925$
nodes and
$783\,027\,125$
undirected links\,\footnote{
We consider here {\em undirected} graphs, see the introduction of
Section~\ref{sec_analysis}.}
with the time (in minutes) at which they were discovered.

\subsubsection{The \ptp\ data-set.}

Several recent studies use traces of
running peer-to-peer file exchange systems to give evidence of some
of their properties, and then design efficient protocols, see for instance
\cite{lefessant04exploiting,voulgaris04exploiting,iptps05}.
They often focus on user behaviors or data properties, and
the complex network approach has proved to be relevant in this context.
Collecting such data however is particularly challenging because of
the distributed and dynamic nature of these systems. Several approaches
exist to obtain data on these exchanges, among which the capture of the
queries processed by a server in a semi-centralized system.

We used here data obtained this way: it contains all the queries
processed by a large {\em eDonkey} server running the {\em Lugdunum}
software \cite{lugdunumurl}. The trace begins from a reboot of the server, on
the 8-th of may, 2004, and lasts until the 10-th, leading to more
than $47$ hours of capture with a time precision of $1$ second.
During this period, the server processed $215\,135\,419$
user commands (logins, logouts and search queries). Here, we
kept the search queries, of the following form:
$T\ Q\ F\ S_1\ S_2\ \dots\ S_n$, where $T$ is the time at which this
query was treated, $Q$ is the peer which sent this query, $F$ is
the queried file, and $S_1$, $S_2$, $\dots$, $S_n$ is a list of
possible providers for this file (they declared to the server that
they have it) sent to $Q$ by the server (so that $Q$ may contact
them directly to get the file).
The trace contains $212\,086\,691$ such queries.

We constructed the {\em exchange graph}, obtained from this data
by considering that, for
each query, at time $T$,
the links between $Q$ and $S_i$ appear for each $i$. This graph
captures some information on exchanges between peers, which is
commonly used as a reasonable image of actual exchanges, see for
instance \cite{iptps05,leblond2004p2p}. The final exchange graph
we obtained has $5\,792\,297$ nodes and $142\,038\,401$ links.

\subsubsection{The \ip\ data-set.}

Since a few years, it has appeared clearer and clearer that measuring
the way computer networks (and their users) behave in running
environments is essential. This is particularly true for the 
internet, where very little is known on large-scale phenomena
like end-to-end traffic or anomalies (congestions, failures,
attacks, etc). In this spirit, several projects measure and
study internet traffic, see for instance \cite{lakhina05imc,lakhina05performance,metrosecurl}.

Here we obtained from the {\em MetroSec} project \cite{metrosecurl}
the following
kind of traces. They record the headers of all IP packets
managed by some routers during the capture period of time.
The trace we use here consists in
a capture done on the router at the interface between
a large laboratory \cite{laasurl} and the outside internet, 
between March 7-th, 08:10~am,  and March 15-th, 2006, 02:22~pm,
leading to a trace of a little more than 8 days and $709\,270\,078$
recorded IP headers.
The trace contains the time at which the packet was managed by
the router, with a precision of $10^{-6}$ second.

From this trace, we extracted for each IP header the sender
and target of the packet, together with the time at which
this packet was routed. We thus obtained the graph
in which nodes are IP addresses and each link represents
the fact that the corresponding IP addresses exchanged (at
least) one packet. Such graphs are used (often implicitely) to study the
properties of exchanges, to seek attack traces, etc. See
for instance \cite{lakhina05imc}.
The final graph used here has $2\,250\,498$ nodes and $19\,394\,216$
links.

\medskip
\section{Analysis}
\label{sec_analysis}

In this section, we present our results on the data-sets described
above. Our aim is to span the main basic properties classically
observed on real-world complex networks. For each set of properties
we recall the appropriate definitions, we discuss their
computation and we analyze their evolution with the size
of the sample in each of our four cases. The key point is that
we compare these behaviors to the classical assumptions in
the field.

In all the definitions in this section, we suppose that a graph
$G=(V,E)$ is given, and we denote by $n = |V|$ its number of nodes,
by $m = |E|$ its number of links, and by $N(v) = \{ u \in V,\ (v,u)\in E \}$
the set of neighbors, or neighborhood, of node $v$. We consider here
undirected graphs (we make no distinction between $(u,v)$ and
$(v,u)$) since most classical properties are defined on
such graphs only. Moreover, recall that our graphs have no loop
and no multiple links, see Section~\ref{sec_approach}.

In order to give precise space and time complexities, we need
to make explicit how we will store our graphs in central memory. We
will use the sorted adjacency arrays encoding: for
each $v \in V$ we store $N(v)$ in a sorted array,
together with its size $|N(v)|$, and access to these
informations is granted in $\Theta(1)$ time and space.
This encoding ensures that the graph is stored in space
$\Theta(m)$ and that the presence of any link can be tested in
$\Theta(\log(n))$ time and $\Theta(1)$ space.

\medskip
\subsection{Basic facts.}
\label{sec_size}
\label{sec_connexity}
\label{sec_basics}

\subsubsection{Size evolution during time.}

As already discussed, in all the paper the properties we consider
will be observed as functions of the sample size, which is the
classical parameter in complex network studies. However, it would also be
relevant to discuss the evolution of these properties during
time\,\footnote{This would reflect the evolution of
the properties during the measurement, not the dynamics of the
complex network under
concern as in \cite{dorogovtsev2003handbook,leskovecdensification}.}.
The plots in Figure~\ref{fig_nodes_links} give the relation between
the two.

\plots{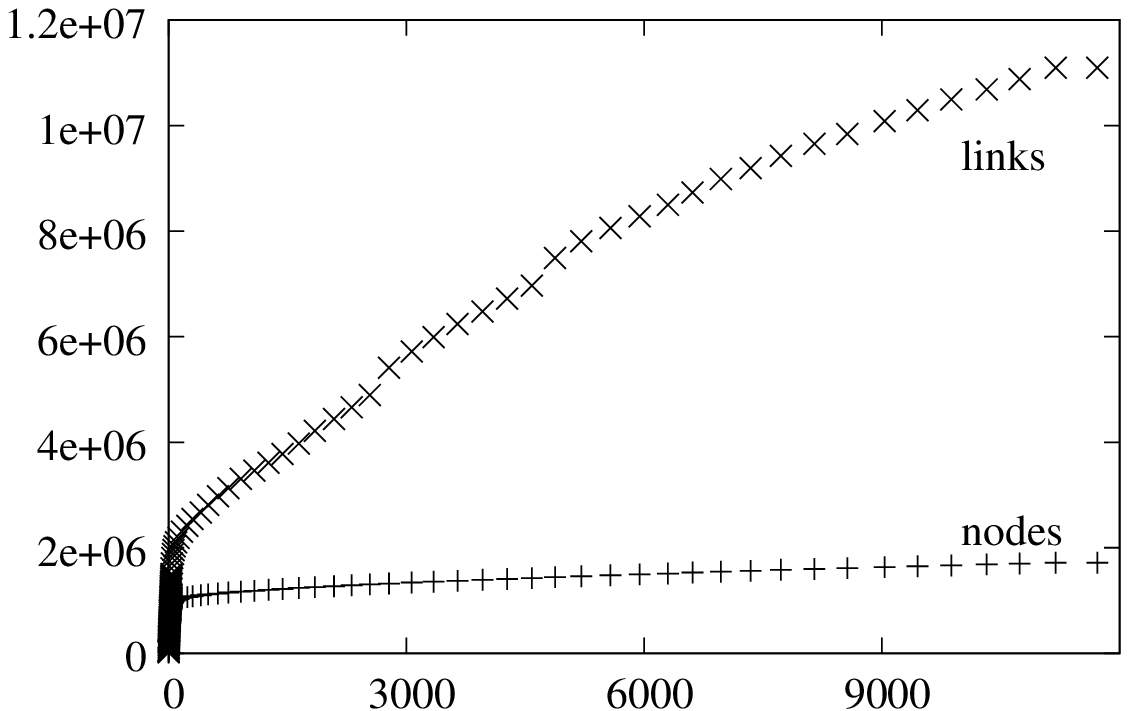}{
Evolution of the number of nodes and links during time (in hours).
}

It appears clearly on these plots that in none of the four cases
does the
measurement reach a state where it discovers no or few
new nodes and links. Instead, the size of the obtained sample
is still growing significantly by the end of the measurement. This
means that, even for huge measurements like the ones we consider,
the final result probably is far from a complete view of the
network under concern. In other words, it is not possible to
collect complete data on these networks in reasonable
time and space, at least using such measurements.

This implies that the observed properties are those of
the samples, and may be different from the ones of the whole
network even at the end of the measurement.
To this regard, an important issue of this contribution
is to determine
whether this is the case or not, and more precisely,
if used samples are representative of
what one would obtain with larger samples or not.

Another important observation is that, in all cases, the number
of links $m$ grows significantly faster than the number of nodes
$n$. We will deepen this in Section~\ref{sec_density}.

Finally, notice that in the case of \inet\ the measurement discovers
a huge number of nodes and links (roughly half the nodes discovered at
the end of the measurement) very quickly.
This is due to the measurement method (based on \traceroute-like
probes) and should not be considered as a surprising fact (it
corresponds to the first probe from each source to each destination).
This will
have an influence on the plots in the rest of the paper: the first half
of each plot will correspond to a very short measurement time. One
may notice that many studies rely on measurement that do only one
probe per destination, thus leading to samples which may be compared
to the ones in the first halves of our plots. However,
as already explained, discussing
this is out of the scope of this contribution.

\subsubsection{Connectivity.}
\label{sec_connectivity}

A connected component of a graph is a maximal (no node can be
added) set of nodes such that a path exists between any pair
of nodes in this set. The connected components and their sizes
are computed using a graph traversal (like a breadth-first
search) in $\Theta(n)$ space and $\Theta(m)$ time.

In most real-world complex networks, it has been observed that
there is a huge connected component, often called
{\em giant component}, together with a number of small
components containing no more than a few percents of the nodes,
often much less, if any.

\plots{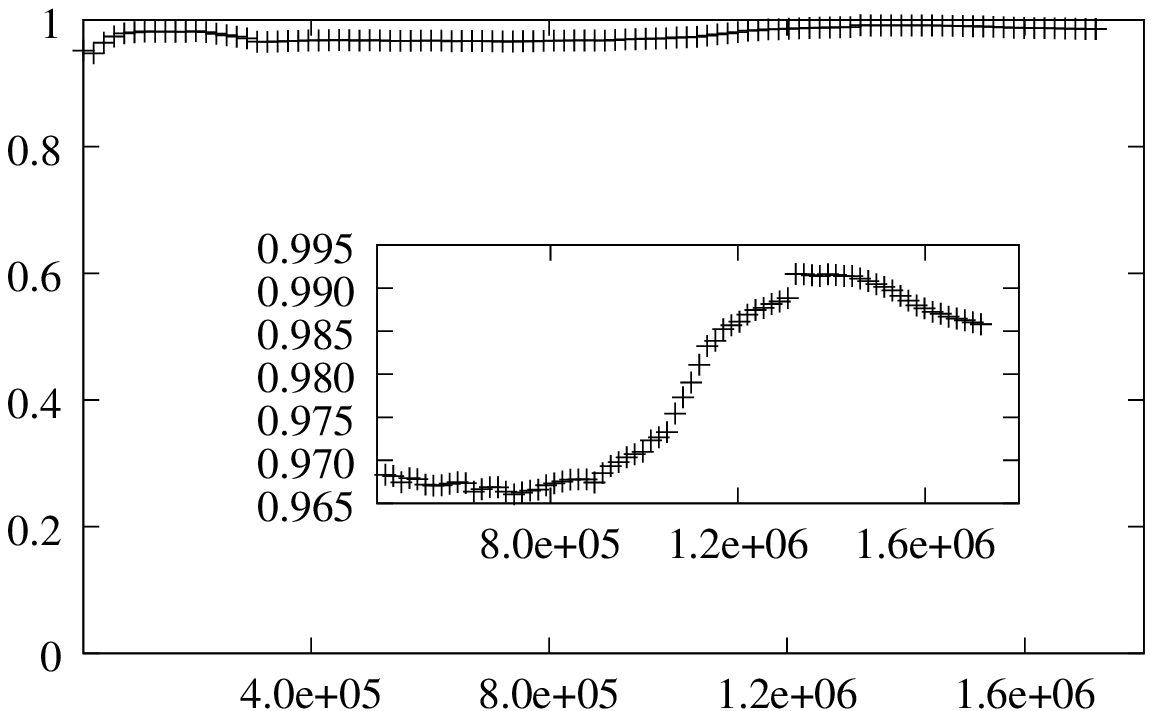}{
Fraction of nodes in the largest connected component
as a function of the sample size, with an inset zoom on
the last three quarters of each plot.
}

\plots{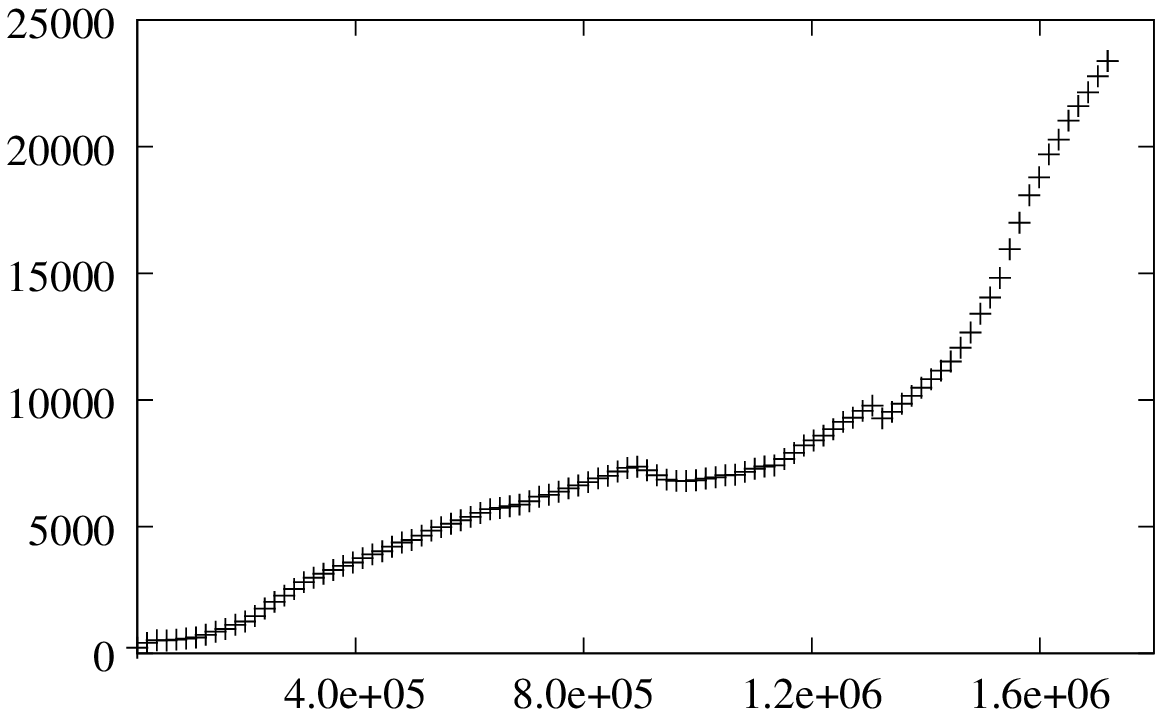}{
Number of connected components
as a function of the sample size.
}

\medskip

In the four cases studied here, these observations are confirmed,
and this is very stable independently of the size of the sample.
This is visible in Figure~\ref{fig_ratio_largest_comp} where
we plot
the proportion of nodes in the giant component: it is very close
to $1$ in all the cases, even for quite small samples (the only
noticable thing is that up to $7\,\%$ of the nodes in \ptp\ are
not in the giant component, but it still contains more than
$92\,\%$ of them). On the
contrary, the number of connected components varies depending
on the case, as
well as its behavior as a function of the size of the graph,
see Figure~\ref{fig_nb_comp}.
Since there is no classical assumption concerning this, and
no clear general behavior, we do not detail these results here.

\medskip
\subsection{Average degree and density.}
\label{sec_avgerage_degree_density}
\label{sec_average_degree}
\label{sec_density}

The degree $d^o(v)$ of a node $v$ is its number of links, or,
equivalently, its number of neighbors: $d^o(v)=|N(v)|$.
The average degree $d^o$ of a graph is the average over all its
nodes: $d^o = \frac{1}{n} \sum_v d^o(v)$. The density is the
number of links in the graph divided by the total number
of possible links: $\delta = \frac{2\cdot m}{n\cdot (n-1)}$. The
density indicates up to what extent the graph is fully connected
(all the links exist). Equivalently, it gives the probability
that two randomly chosen nodes are linked in the graph.
There is a trivial relation between the average degree and the
density: $d^o = \delta \cdot (n-1)$.
Both the average degree and the density are computed in
$\Theta(n)$ time and $\Theta(1)$ space.

The average degree of complex networks is supposed to be small,
and independent of the sample size, as soon as the sample is large enough.
This implies that the density $\delta$ is supposed to go to zero when the sample
grows, since $\delta = \frac{d^o}{n-1}$.

\medskip

It appears in Figures~\ref{fig_avg_deg} and~\ref{fig_density}
that the average degree is indeed
very small compared to its maximal possible value, and that the
density is close to zero, as expected.

\plots{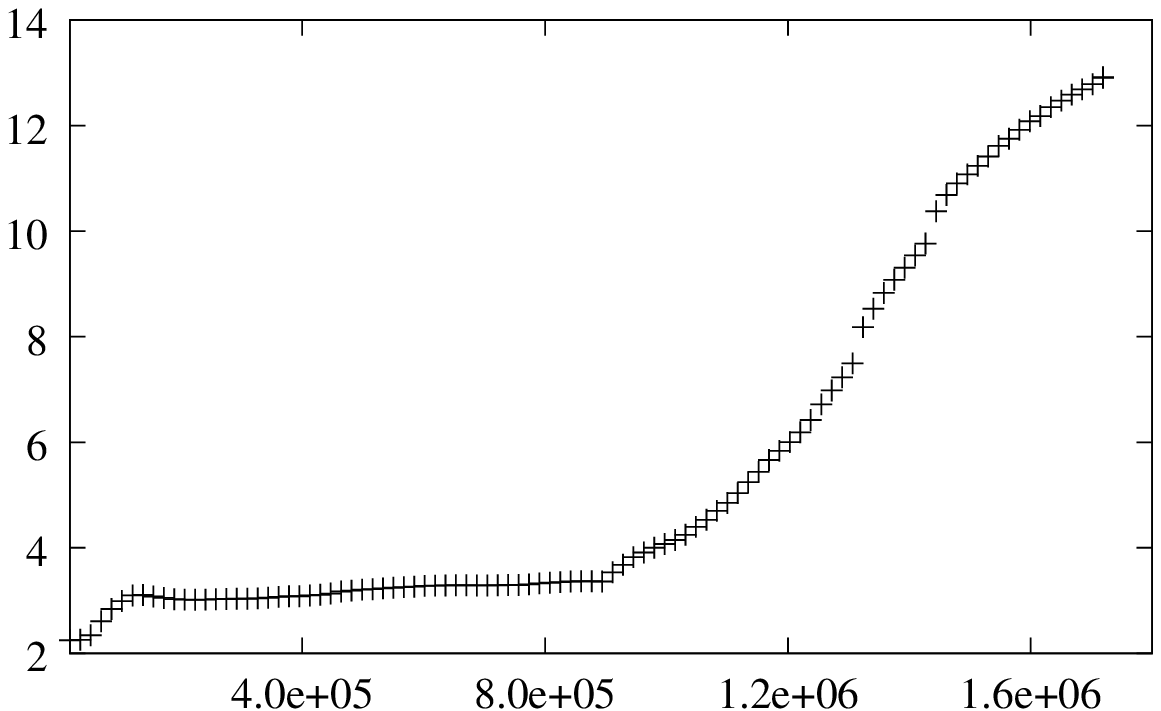}{
Average degree
as a function of the sample size.
}

\plots{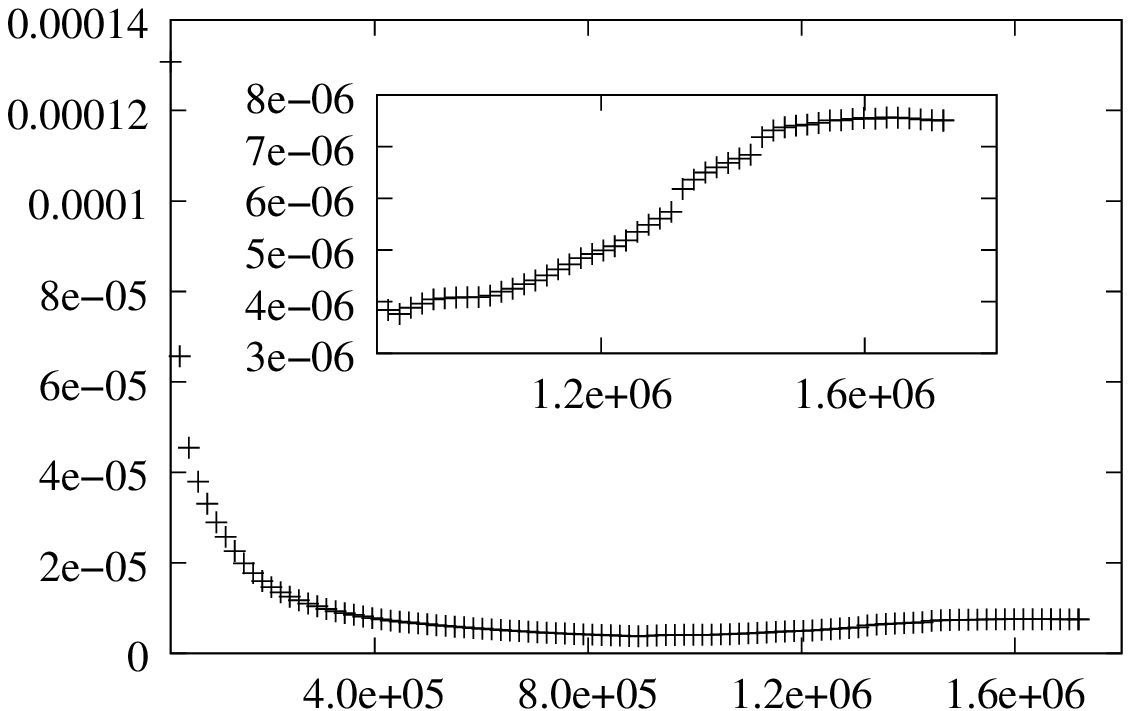}{
Density as a function of the sample size, together with
inset zooms of the rightmost halves of the plots.
}

In the cases of \web\ and \ip, the measurement reaches a regime
in which the average degree is rather stable (around $40$ and $17$,
respectively), and
equivalently the density goes to $0$. This means that there is
little chance that this value will evolve if the sample grows
any further, and that the observed value would be the same
independently of the sample size (as long as it is not too
small). In this sense, the observed value may be trusted, and
at least it is not representative of only one particular
sample. We will discuss this further in Section~\ref{sec_conclusion}.

In the two the other cases, \inet\ and \ptp, the observed average degree is far from
constant, and the density does not go to zero.
This has a strong meaning: in these cases, one {\em cannot}
consider the value observed for the average degree on any
sample as significant. Indeed, taking a smaller or a larger sample
would lead to a different value. Since the measurements we
use here are already huge, this even means that there is
little chance that the observed value will reach a steady
state within reasonable time using such measurements.
We will discuss this further in Section~\ref{sec_conclusion}.

\medskip

Going further, one may observe that in some cases the number
of links $m$ grows faster than the number of nodes $n$ (the
average degree grows), and even
as $n^2$ (the density is stable) in some parts of the plots.
In order to deepen this, we present the plots of $m$ as
a function of $n$ in Figure~\ref{fig_links_f_nodes}, in
log-log scales: straight lines indicate that $m$ evolves
as a power of $n$, the exponent being the slope of the
line.

\plots{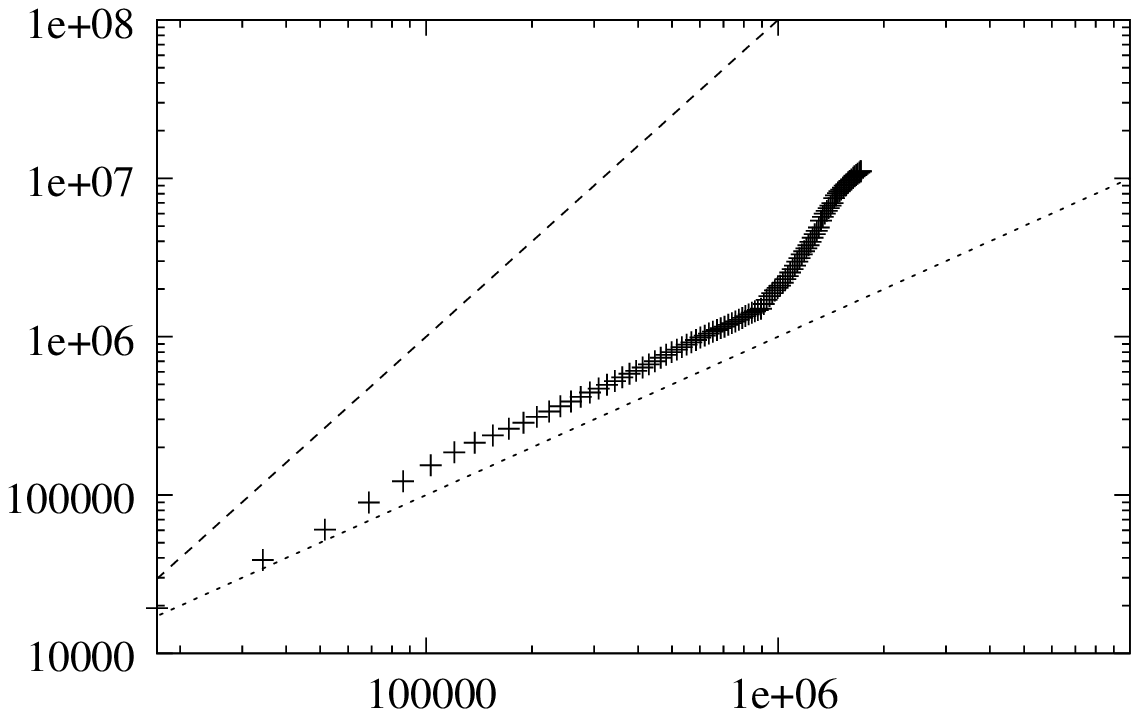}{
Number of links as a function of the number of nodes
in log-log scales,
together with the plots of $y=x$ and $y=x^2$ (with an
appropriate shift).
}

Such plots have been studied in the context of dynamic graphs
\cite{leskovecdensification}. In
this paper, the authors observe that $m$ seems to evolve
as a power of $n$, and that the average degree grows with
time, which was also observed in \cite{dorogovtsev2003handbook}.
In our context, the behavior of $m$ as a function of $n$ is
quite different: the plots in Figure~\ref{fig_links_f_nodes}
are far from straight lines in most cases. This means that
exploring more precisely the relations between $m$ and $n$
needs significantly more work, which is out of the scope of
this paper. The key point here is that, in some cases, $m$
grows faster than $n$, and that the classical algorithmic
assumption that $m \in \Theta(n)$ is not always true.

\medskip

Finally, the properties observed in this section are in sharp
contradiction with the classical assumptions of the field for
two of our four real-world cases (\inet\ and \ptp).
This means that, in these cases, one cannot assume that the
average degree observed with such a measurement is representative
of the one of the actual network: taking a larger or smaller
sample leads to significantly different estimations.
In the two other cases (\web\ and \ip), instead, the measurement
seems to reach a state where the observed values are
significant.

\medskip
\subsection{Average distance and diameter.}
\label{sec_average_distance}
\label{sec_diameter}
\label{sec_distances}

We denote by $d(u,v)$ the distance between $u$ and $v$, \ie\ the number
of links on a shortest path between them.
We denote by $d(u) = \frac{1}{n} \sum_v d(u,v)$ the average distance from
$u$ to all nodes, and by
$d = \frac{1}{n} \sum_u d(u) = \frac{1}{n^2} \sum_{u,v} d(u,v)$ the
average distance in the considered graph. We also denote by
$D = \max_{u,v} d(u,v)$ the diameter of the graph, \ie\ the largest
distance.

Notice that the definitions above make sense only for connected graphs.
In practice, one generally restricts the computations to the largest
connected component, which is reasonable since the vast majority of
nodes are in this component (see Section~\ref{sec_connectivity}). We
will follow this convention here; therefore, in the rest of this subsection,
the graph is supposed to be connected (\ie\ it has only one connected
component) and the computations are made only on the giant component of
our graphs.

\subsubsection{Computation.}

Computing distances from one node to all the others in an undirected
unweighted graph can be done in $\Theta(m)$ time and $\Theta(n)$ space
with a breadth-first search (BFS).
One then obtains all the distances in the graph, needed for exact
average distance and diameter computations, in $\Theta(n\cdot m)$
time and $\Theta(n)$ space. This is space efficient, but not fast
enough for our purpose (see Section~\ref{sec_approach}).
Faster algorithms have been proposed \cite{alon1992boolean,seidel1992apsp,feder1991clique},
but they all have a $\Theta(n^2)$ space cost, which is
prohibitive in our context. See \cite{zwick01exact} for a survey,
and \cite{palmer2002tool,eppstein2004centrality} for recent results on the topic.

Despite this, the average distance and the diameter are among the most
classical properties used to describe real-world complex networks.
Therefore, computing accurate estimations of the average distance and
the diameter is needed, and much work has already be done to this
regard \cite{zwick01exact,palmer2002tool,eppstein2004centrality}.

A classical approach is to approximate the average distance
by using a limited number of BFS and then average over this sample.
See \cite{eppstein2004centrality} for formal results on this.
We used here a variant of this approach:
at step $i$ we choose a random node, say $v_i$, and we compute its
average distance to all other nodes, $d(v_i)$, in time $\Theta(m)$ and
space $\Theta(n)$.
Then 
we compute the $i$-th approximation of the average distance as
$d_i = \frac{1}{i} \sum_{j=1}^i d(v_j)$. The loop ends at the first
$i>i_{\min}$ such that
the variations in the estimations
have been less than $\epsilon$ during the last $i_{\min}$ steps,
{\em i.e.}
$|d_{j+1}-d_j| < \epsilon$,
for all $j, i-i_{\min} \le j < i$.
The variables
$i_{\min}$ and $\epsilon$ are parameters used to ensure that at least
$i_{\min}$ iterations are processed, and that the variation during the
$i_{\min}$ last iterations is no more than $\epsilon$. In all the
computations below, we took $i_{\min} = 10$ and $\epsilon = 0.1$.

\medskip

Such approaches are much less relevant for notions like the diameter,
which is a worst case notion: by computing the worst case on a
sample, one may miss a  significantly worse case. Instead, we propose
simple and efficient algorithms to find lower and upper bounds for
the diameter.

First notice that the diameter of a graph is at least the height of
any BFS tree of this graph. Going further,
it is shown in \cite{corneil2003diameter,CDHP01} that the following
algorithm finds excellent approximations of the diameter of graphs
in some specific cases: given a randomly chosen node
$v$, one first finds the node $u$ which is the further from $v$ using
a BFS, and then processes a new BFS from $u$; then the lower bound
obtained from $u$ is at least as good as the one obtained from $v$,
and is very close to the diameter for some graph classes.

Now, notice that the diameter of a graph cannot be larger than
the diameter of any of its (connected) subgraphs, in particular
of its BFS trees. Therefore the diameter is bounded by the
largest distance in any of its BFS trees, which can be computed
in $\Theta(n)$ time and space, once the BFS tree is given. One
then obtains an upper bound for the diameter in the graph.

We finally iterate the following to find accurate
bounds for the diameter.
Randomly choose a node and use it to find a lower bound using the algorithm
described above; then choose a node in decreasing order of degrees and use
it to find an upper bound as described above. In the latter, nodes
are chosen in decreasing order of their degrees because high degree nodes
intuitively lead to BFS trees
with smaller diameter. We iterate this at least $10$ times, and until
the difference between the two bounds becomes lower
than $5$. In the vast majority of the cases considered here, the $10$
initial steps are sufficient.
Since each step needs only $\Theta(m)$ time and $\Theta(n)$ space,
the overall algorithm performs very well in our context.

\subsubsection{Usual assumptions and results.}

It appeared in various contexts (see for instance
\cite{watts1998smallworld,kleinberg99web,albert99nature,broder00graph})
that the average distance and the diameter of real-world
complex networks is much lower than expected, leading to the
so-called {\em small-world} effect: any pair of nodes tends to be connected
by very short paths. Going further,
both quantities are also supposed to grow slowly with the number of
nodes $n$ in the graph (like its logarithm or even slower).

\plots{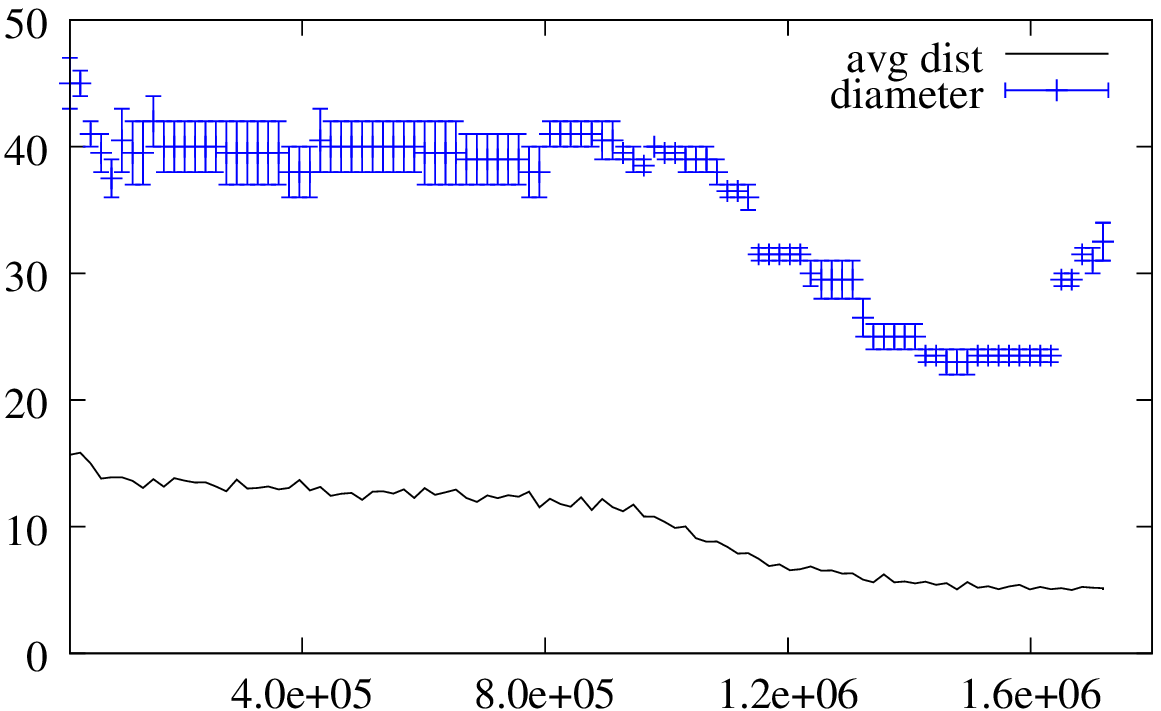}{
Estimation of the average distance and bounds for the diameter,
as a function of the sample size.
}

\medskip

Figure~\ref{fig_distances} shows several things. First, the obtained
bounds for the diameter are very tight and give a precise information
on its actual value. The heuristics described above therefore are
very efficient and provide a good alternative to previous methods in
our context.
These plots also indicate that our approximation of the average
distance is consistent: if the randomly chosen nodes had a significant
impact on our evaluation, then the corresponding plots would not be
smooth.

Concerning the obtained values themselves, they clearly confirm that
both the average distance and the diameter are very small compared
to the size of the graphs. However, their evolution is in sharp
contrast with the usual assumptions in the case of \inet\ and \ip:
both the average distance and the diameter are stable or even
decrease\,\footnote{Similar behaviors were observed
in \cite{leskovecdensification} in the context of dynamic graphs,
leading to the claim that these graphs have {\em shrinking diameters}.}
with the size of the sample in these cases (with a sharp increase at the
end for the diameter of \inet).
In the case of \web, however, the observed behavior fits very
well classical assumptions. The situation is not so clear for \ip:
the values seem stable, but they may grow very slowly.

These surprising observations may have a simple explanation. 
Indeed, the usual assumptions concerning average distance and
diameter are strongly supported by the fact that the average
distance and diameter of various random graphs (used to model
complex networks) grow with their size.
However, in these models, the average degree $d^o$ generally is supposed to
be a constant independent of the size. If it is not, then the
average distance in these graphs typically grows as $\frac{\log(n)}{\log(d^o)}$
\cite{bollobas01random,newman01arbitrary}.
This means that, if $d^o$ grows with $n$ as observed in
Section~\ref{sec_density}, it is not surprising that the average
distance and the diameter are stable or decrease. Likewise, in the
case of \web\ where the average degree is constant, the average
distance and the diameter should increase slowly, which is in
accordance with our observations.

\medskip
\subsection{Degree distribution.}
\label{sec_degrees}
\label{sec_deg_distrib}

The degree distribution of a graph is the proportion $p_k$ of nodes
of degree exactly $k$ in the graph, for all $k$. Given the encoding
we use, its computation is in $\Theta(n)$ time and space.

Degree distributions
may be homogeneous (all the values are close to the average, like
in Poisson and Gaussian distributions), or heterogeneous (there is
a huge variability between degrees, with several orders of magnitude
between them). When a distribution is heterogeneous, it makes sense
to try to measure this heterogeneity rather than the average value.
In some cases, this can be done by fitting the distribution by a
power-law, \ie\ a distribution of the form $p_k \sim k^{-\alpha}$.
In such cases, the exponent $\alpha$ may be considered
as an indicator of how heterogeneous the distribution is.

\subsubsection{Usual assumptions and results.}

Degree distributions of complex networks have been identified as a
key property since they are very different from what was thought until
recently \cite{faloutsos99sigcomm,kleinberg99web}, and since it
was proved that they have a crucial
impact on phenomena of high interest like network robustness \cite{albert00error,kim2004random}
or diffusion processes \cite{pastor01epidemic,ganesh2005epidemics}. They are considered to be highly
heterogeneous, generally well fitted by a power-law, and independent
of the size of the graph.

We first present in Figure~\ref{fig_final_dd} the degree distributions
observed in our four cases at the end of the measurement procedure.
These plots
confirm that the degrees are very heterogeneous, with most nodes
having a low degree ($49\%$, $39\%$, $24\%$ and $93\%$ have degree
lower than $5$ in \inet, \ptp, \web\ and \ip\ respectively), but some
nodes having a very high degree (up to $35\,455$, $15\,115$, $1\,776\,858$ and
$259\,905$ in \inet, \ptp, \web\ and \ip).
We however note that the \ptp\ degree distribution
does not have a heavy tail, but rather an exponential cutoff.
All the degree distributions are reasonably, but not perfectly,
fitted by power laws on several decades.

\plots{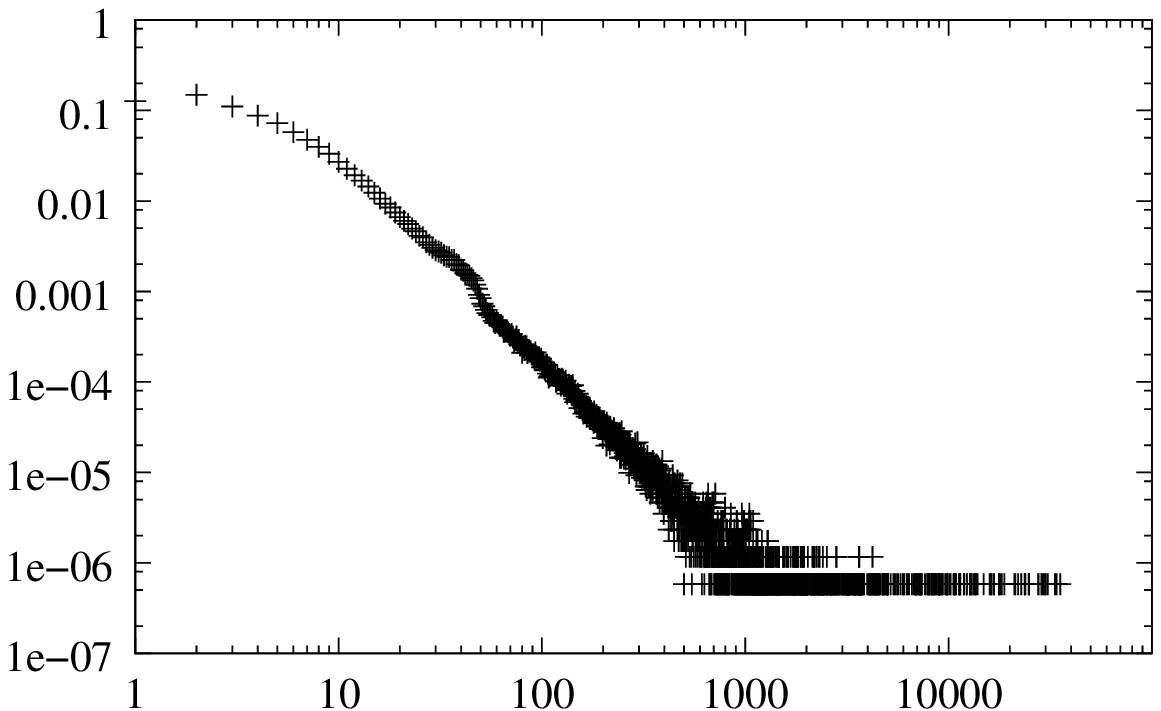}{
Degree distributions of the final samples.
}

But recall that our aim is to study how the degree distribution
{\em evolves} when the size of the sample grows. In order to
do this, we will first plot
cumulative distributions (\ie\ for all $k$ the proportion
$q_k = \sum_{i\ge k} p_i$ of nodes of degree at least $k$),
which are much easier to compare
empirically than actual distributions.
In Figure~\ref{fig_trois_distr_cum} we show the cumulative distributions
in our four cases, with three different sample sizes each. These
plots show that the fact that the degrees are highly heterogeneous
does not depend on the sample size: this is true in all cases.

\plots{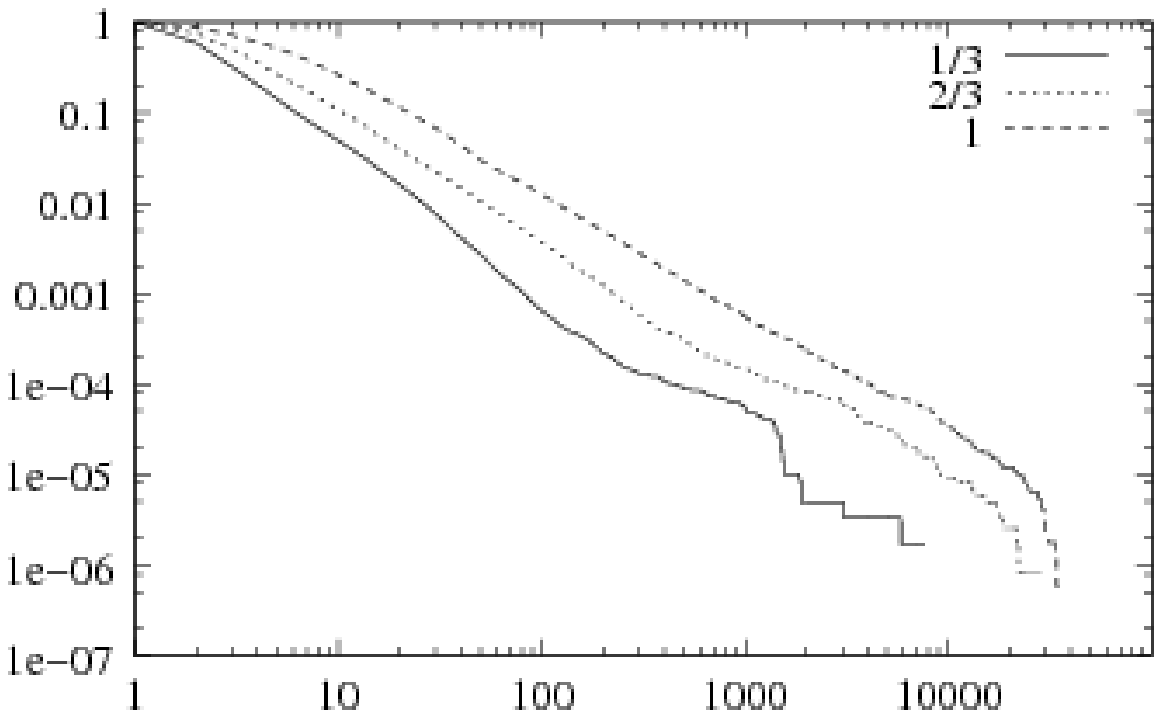}{
Cumulative degree distributions for different sample sizes
($\frac{1}{3}$ and $\frac{2}{3}$ of the total, and the
total itself).
}

One may however observe that for \inet\ and \ip\ 
the distributions significantly change as the samples grow.
In the \inet\  case one may even be tempted to say that the
{\em slope}, and
thus the exponent of the power-law fit, evolves.
We will however avoid such conclusions here: the difference is not
significant enough to be observed this way.

In the case of \web, only the maximal degree significantly changes.
Notice that, in this case, the average degree is roughly constant,
meaning that this change in the maximal degree has little impact
on the average. This is due to the fact that it concerns only very
few nodes. In the case of \ip, the changes are mostly between the
values $10$ and $200$ of the degree; below and above this interval,
the distribution is very stable, and even there the global shape
changes only a little.

\medskip

At this point, it is important to notice that the fact that
the degree distributions evolve (for \inet\ and \ptp) is not surprising,
since the average degree itself evolves, see
Section~\ref{sec_density}.
In order to deepen this, we need a way to quantify the difference
between degree distributions, so that
we may observe their evolution more precisely.

The most efficient way to do so probably is to use the classical
Kolmogorov-Smirnof (K-S) statistical test, or a similar one.
Given two distributions $p_k$ and $p'_k$ which we want to compare,
it consists in computing the maximal difference $\max_k(|q_k-q'_k|)$
between their respective cumulative distributions $q_k$ and $q'_k$.
This test is known to be especially well suited to compare heterogeneous
distributions, when one wants to keep the comparison simple.

We display in Figure~\ref{fig_k_s_test_final} the values obtained by
the K-S test when one compares the degree distribution at each step
of the measurement to the final one. This makes it
possible to see how the degree distribution evolves towards the final
one as the sample size grows.

The K-S test may first have a phase where it varies much
but finally reach a phase where its value
oscillates close to $0$ (note that it cannot be negative),
indicating that the measurement
reached a stable view of the degree distribution. This is what
we observe in the \web\ and \ip\ cases, confirming the fact that the
degree distribution is very stable in these cases
(Figure~\ref{fig_trois_distr_cum}).
However, the K-S test has a totally different behavior in the
other cases: it shows that the degree distribution
continuously varies during the measurement. This means that its
observation on a particular sample cannot be considered as
representative in these cases.
We will discuss this further in Section~\ref{sec_conclusion}.

\plots{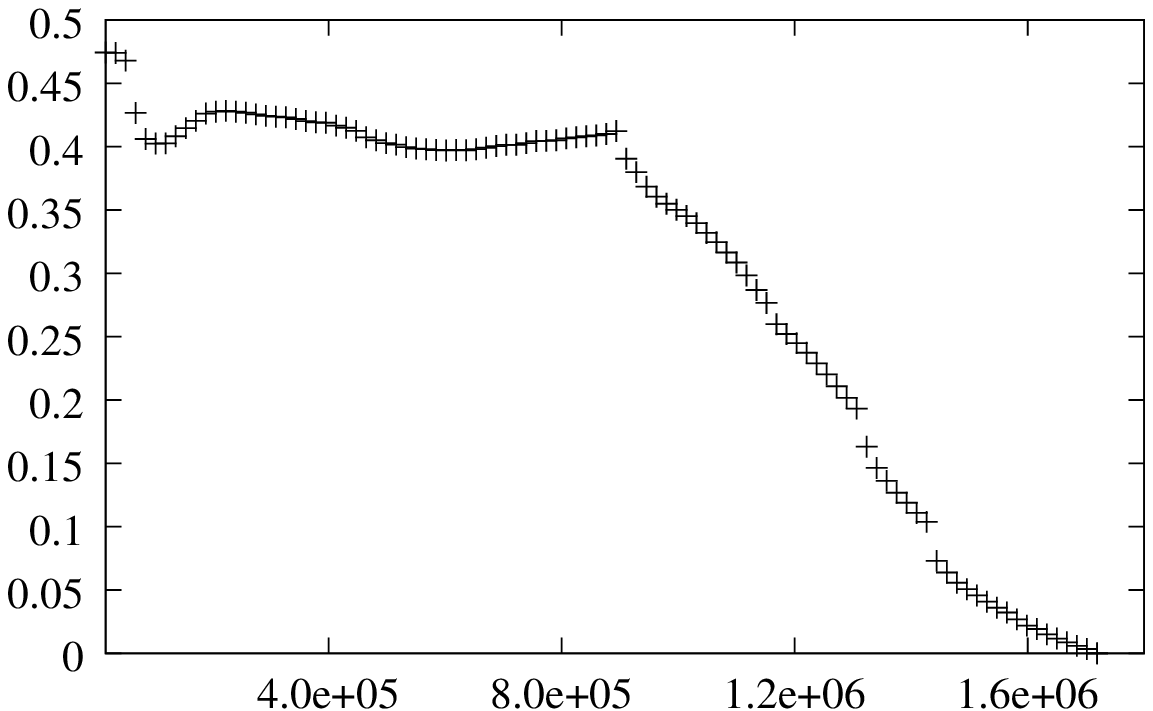}{
Evolution of the degree distribution according to a K-S test with the final one,
as a function of the sample size.
}

Going further, notice that, in several cases, the evolution of the K-S
test is strongly related to the one of the average degree, see
Figures~\ref{fig_avg_deg} and~\ref{fig_k_s_test_final}: the plots are
almost symmetrical for \inet\ and \web, and in the two other cases
there also seems to be a strong relation between the two statistics.
However, there exist cases where their behaviors are very different,
which may be observed here for instance for small sizes of the
\ip\ samples. This confirms that the K-S
test captures other information than simply the average degree,
and therefore the similarities observed here are nontrivial:
here, the evolution of the degree distributions is well
captured by the evolution of the average degree itself, as long
as the sample is large enough. In other
words, when the average degree does not change, the KS-test
(and thus the main properties of the degree distribution) also is
stable, in our cases.

\medskip

Let us finally notice that methods exist to automatically compute
the best
power-law fit of a distribution according to various criteria.
The simplest one probably is a least-square linear fit of the
log-log plot, but it can be improved in several ways and more
subtle methods exist, see for instance~\cite{goldstein2004fitpl,newman2005powerlaw}.
Such automatic approaches are appealing in our context since
they would allow us to plot the evolution of the exponent of the
best fit as a function of the sample size.

We tried several such methods, but it appears that our degree
distributions are too far from perfect power-laws to give significant
results. We tried both with the classical distributions and the
cumulative ones, and both with the entire distributions and
with parts of them more likely to be well fitted by
power-laws. The results remain poor, and vary depending
on the used approach (including the fitting method). We
therefore consider them as not significant, and we do not
present them here.

\medskip
\subsection{Clustering and transitivity.}
\label{sec_clustering}

Despite having a small density, a graph may have a high {\em local} density: if two nodes
are close to each other in the graph, they are linked together with a much higher
probability than two randomly chosen nodes. There is a variety of ways to capture
this, the most widely used being to compute the clustering coefficient and/or the
transitivity ratio, which we will study in this section.

The clustering coefficient of a node $v$ (of degree at least $2$) is the probability
for any two neighbors of $v$ to be linked together:
$\cc(v) = \frac{2\cdot|E_{N(v)}|}{d^o(v)\cdot(d^o(v)-1)}$
where $E_{N(v)} = E \cap (N(v)\times N(v))$ is the set of links between
neighbors of $v$. Notice that it is nothing but the density of the neighborhood
of $v$, and in this sense it captures the local density.
The clustering coefficient of the graph itself is the average of
this value for all the nodes (of degree at least $2$):
$\cc = \frac{1}{|\{v\in V,\ d^o(v)\ge2\}|} \sum_{v\in V,\ d^o(v)\ge2} \cc(v)$.

One may also define the transitivity ratio of the graph as
follows:
$\tr = \frac{3\cdot N_{\Delta}}{N_{\vee}}$
where $N_{\Delta}$ denotes the number of triangles, \ie\ sets of three
nodes with three links, in the graph and $N_{\vee}$ denotes the number  of
connected triples, \ie\ sets of three nodes with two links,
in the graph.

Computing the clustering coefficient and transitivity ratio
is strongly related to counting and/or listing all the
triangles in a graph. These problems have been well studied,
see \cite{triangles06} for a survey.
The fastest known algorithms have a space complexity in
$\Theta(n^2)$, which is prohibitive in our context. Instead,
one generally
uses a simple algorithm that computes the number of triangles
to which each link belongs in $\Theta(n\cdot m)$ time and
$\Theta(1)$ space. This is too slow for our purpose, but
more subtle algorithms exist
with $\Theta(m^{\frac{3}{2}})$ time and
$\Theta(n)$ space costs in addition to the $\Theta(m)$ space
needed to store the graph. Some of them moreover have the advantage
of performing better on graphs with heterogeneous degree
distributions like the ones we consider here, see Section
\ref{sec_deg_distrib}. We use here such an
algorithm, namely {\em compact-forward}, presented in
\cite{schank05wea,triangles06}.

\subsubsection{Usual assumptions and results.}

Concerning clustering coefficients, there are several assumptions
commonly accepted as valid. The key ones are the fact that the
clustering coefficient and the transitivity ratio are significantly
(several orders of magnitude) larger than the density, and that they
are independent of the sample size, as long as it is large enough.
Moreover, the two notions are generally thought as equivalent.

\plots[0.34]{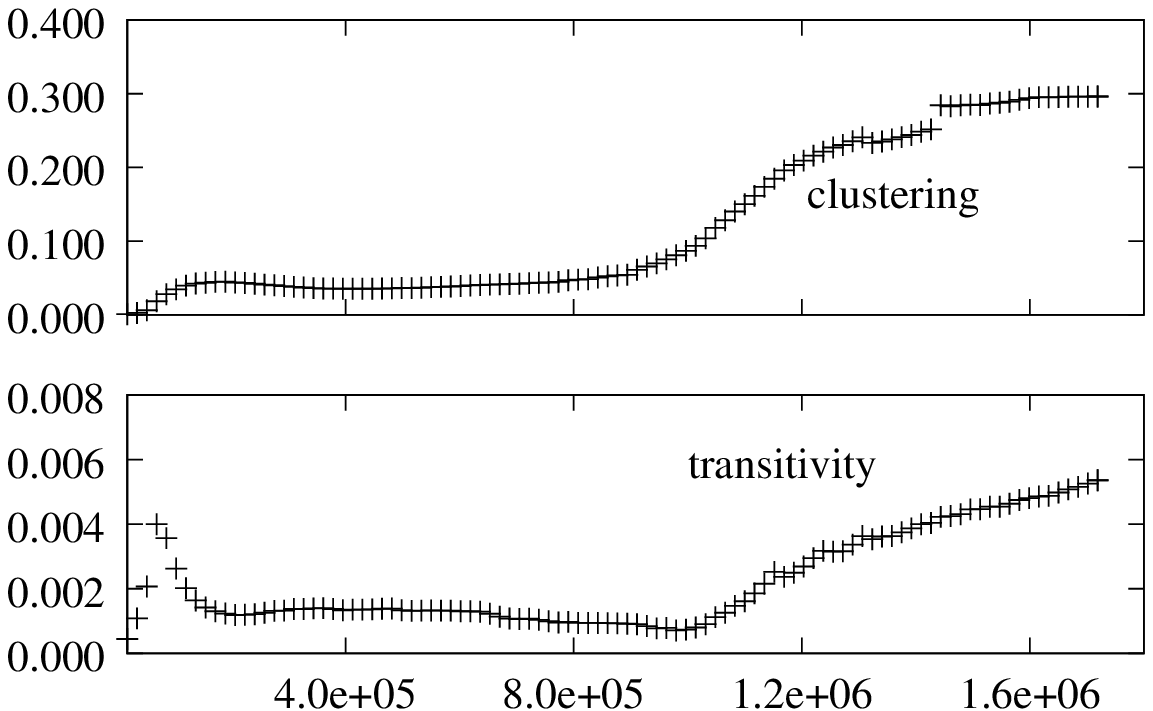}{
The clustering coefficient and transitivity ratio
as a function of the sample size.
}

Let us first notice that, because of its definition (see Section~\ref{sec_data})
the \ip\ graph can contain only very few triangles: most of its links
are between nodes inside the laboratory and nodes in the outside
internet, which prevents triangle formation. Observing
the clustering coefficient and the transitivity ratio on such graphs
makes little sense. Therefore, we will show the plots but we will not
discuss them for this case.

It appears clearly  in Figure~\ref{fig_clustering} that the values of
both statistics are indeed much larger than the density in our examples
(except for \ip, as explained above).
But it also appears that their value is quite
unstable (except in part for \ptp); for instance the transitivity ratio in
the \inet\ graph experiences a variation of approximately $4$ times its own
value. Moreover, the clustering coefficient and the transitivity ratio
evolve quite differently (they even have opposite slopes in the \web\ case).
Finally, there is no general behavior, except that the observed value
is unstable in most cases. This indicates that it is unlikely 
that one may infer the clustering coefficient or the transitivity ratio of the 
underlying complex network from such measurements, and that
the values obtained on a given sample are not representative (except
the transitivity ratio of \ptp, in our cases).
We will discuss this further in Section~\ref{sec_conclusion}.

\medskip

At this point, it is important to notice that for the statistics we
observed previously, each one of our graphs conformed to either all or none of the usual
assumptions. This is not the case anymore when
we take the clustering coefficient and the transitivity ratio into account.
Typically, despite the fact that it conforms to all other classical
assumptions on the properties we studied until now, \web\ does not have
stable values for these new statistics. Conversely, the
transitivity ratio of \ptp\ is very stable whereas its observed
properties did not match usual assumptions until now.
This shows that, while the properties studied in previous sections
seem to be strongly related to the average degree, the ones observed here
are not.

\plots{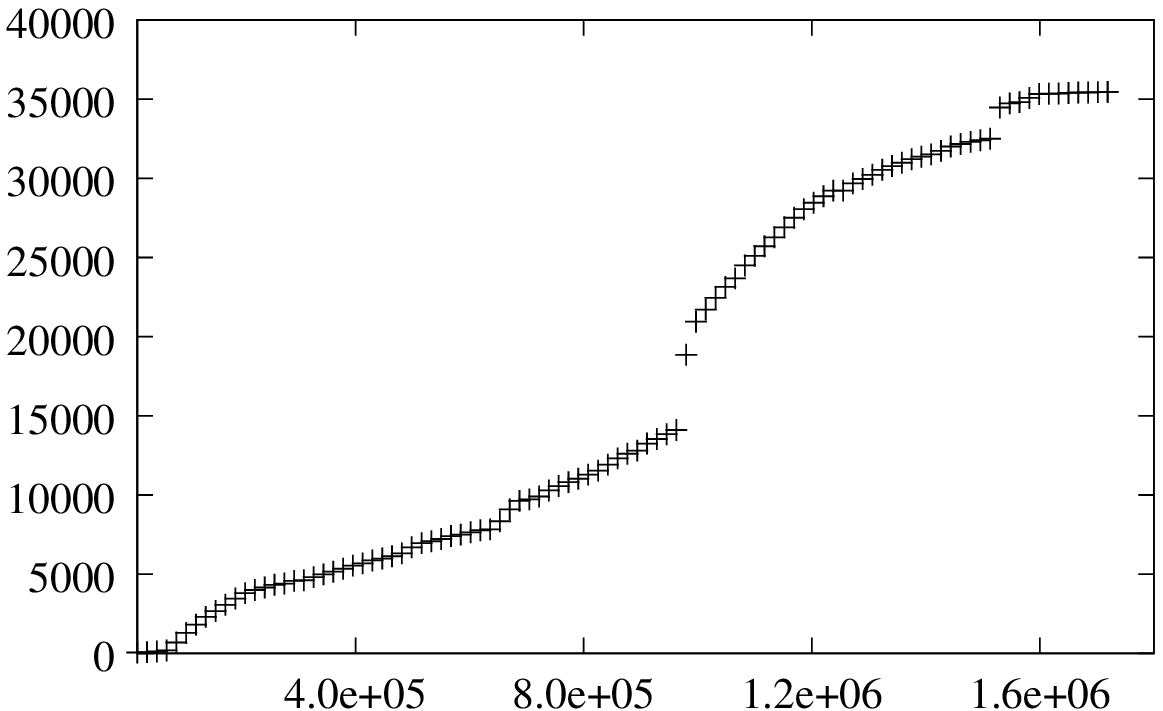}{
Maximal degree as a function of the sample size.
}

\plots{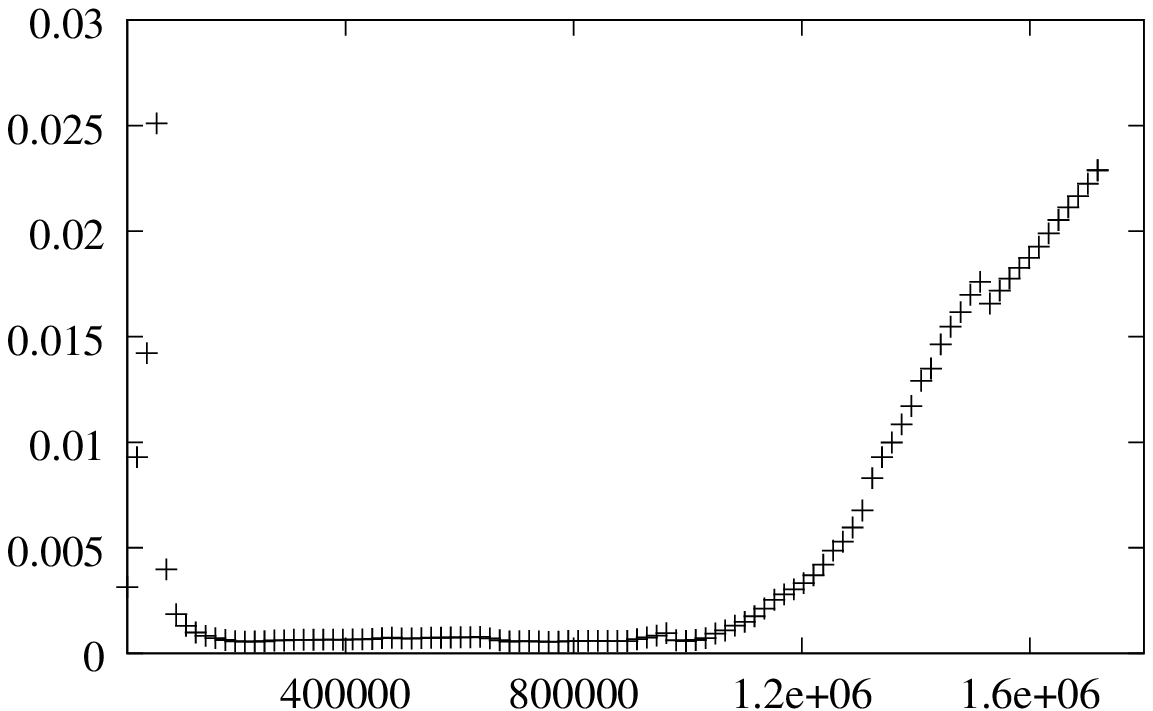}{
Number of triangles divided by the square of the maximal degree,
as a function of the sample size.
}

One may therefore investigate other explanations. We already observed
in Section~\ref{sec_degrees} that, in the case of \web, the maximal
degree is not directly related to the average degree: it varies
significantly though the global distribution and the
average degree are stable. Going further, we plot the maximal degree
$d_{\max}$ of our samples as a function of their size in
Figure~\ref{fig_max_deg}.
It seems that it is correlated to the variations of the
transitivity ratio. This is due to the fact that the maximal degree node
plays a key role in the number of connected triples in the graph:
it induces approximately ${d_{\max}}^2$ such triples. Therefore, any
strong increase
of the maximal degree induces a decrease of the transitivity ratio,
and when the maximal degree remains stable the transitivity ratio tends
to grow or to stay stable\,\footnote{As a consequence, one may
consider that the transitivity ratio is not relevant in graphs where
a few nodes have a huge degree: these nodes dominate the behavior of
this statistics. This has already been
discussed, see for instance \cite{schank04approximating}, but this is
out of the scope of this contribution.}.
This is confirmed by the plot of the number of triangles divided by
the square of the maximal degree, as a function of the sample size,
Figure~\ref{fig_nb_tr_deg_max},
which has a shape similar to the transitivity plots.

\plots{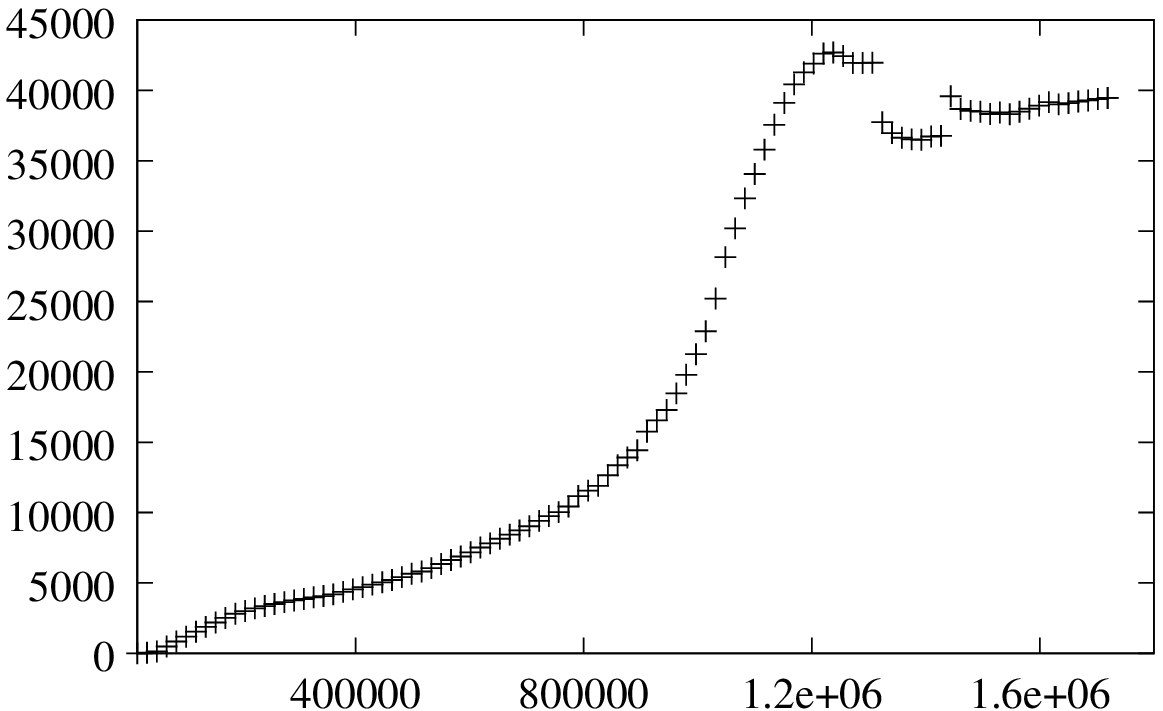}{
Clustering coefficient divided by the density,
as a function of the sample size.
}

Concerning the clustering coefficient, which captures the {\em local}
density, the important points in usual assumptions are that it is
several orders of magnitude larger than the (global) density and that it is
independent of the sample size. Since the second part of this claim
is false, and since the usual assumptions on density are also false, one may
wonder how the ratio between the two values evolves. Figure~\ref{fig_ccloc_dens}
shows that this ratio tends to be constant when the sample becomes
very large, especially for the \ptp\ and \ip\ cases. This is
a striking observation indicating that the ratio between
density and clustering coefficient may be a much more relevant
statistical property than the clustering coefficient in our context:
it would make sense to seek accurate estimations of this ratio using practical
measurements, rather than estimations of the two involved statistics on their own.

\section{Conclusion and discussion.}
\label{sec_conclusion}

In this paper, we propose the first practical method to
rigorously evaluate the relevance of properties observed on large
scale complex network measurements. It consists in studying how these
properties evolve when the sample grows during the measurement.
Complementary to other
contributions to this field \cite{lakhina02sampling,barford01imc,achlioptas05bias,guillaume2005metro,dallasta04statistical}, this method
deals directly with real-world data, which has the key advantage
of leading to practical results.

We applied this methodology to very large measurements of four different
kinds of complex networks. These data-sets are significantly larger than
the ones commonly used, and they are representative of the wide
variety of complex networks studied in computer science.
The classical approach for studying these networks is to collect as much
data as possible (which is limited by computing capabilities and
measurement time, at least), and then to assume that the
obtained sample is representative of the whole.

\medskip

Our key result is that our methodolody makes it possible to rigorously
identify cases where this approach is misleading, whereas in other cases
it makes sense and may lead to accurate estimations.

\medskip

In the case of \inet, for instance, the average degree of the
sample grows with its size (once it is large enough), which
shows clearly that the average degree observed on a particular sample
is certainly not the one of the whole graph.
In the case of \web, on the contrary, the average degree reaches a stable
value, indicating that collecting more data probably would not
change it. Despite this, the transitivity ratio of this
graph is still unstable by the end of the measurement, which shows
that a given measurement may reach a stable regime for some of its basic
properties while others are still unstable. This is confirmed by
\ptp, which has a stable transitivity ratio but unstable average
degree. These last observations also show that there is no clear
hierarchy between properties: the stability or unstability of some
properties are independent of each other.

Some observations we made on these examples are in sharp contrast
with usual assumptions, thus proving that these assumptions are erroneous in
these cases.
Other observations are in accordance with them, which
provides for the first time a rigorous empirical argument
for the relevance of these assumptions in some cases.

More generally, the proposed method makes it possible to distinguish
between the two following cases:
\begin{itemize}
\item either the property of interest does not reach a stable regime
during the measurement, and then
this property observed on a given sample certainly is
erroneous;
\item or the property does reach a stable regime, and then
we may conclude that it will probably not evolve anymore and that it
is indeed a property of the whole network (though it is possibly biased, see below).
\end{itemize}

\medskip

The fact that, even if it is stable, the observed property may be
biased is worth deepening. Indeed, it may actually evolve again when the
sample grows further (like the average degree in
our \inet\ measurement for
instance, see Figure~\ref{fig_avg_deg}). This makes the collection of very large data-sets a key
issue for our methodology.

This does not entirely solve the
problem, however: the property may remain
stable until the sample spans almost all the network
under concern, but still be significantly biased; 
finite-size effects may lead to variations in the observation
at the end of the measurement (like at its beginning). Moreover,
the fact that the underlying network evolves during
the measurement should not be neglected anymore. Going even
further, one may notice that some measurement techniques are
unable to provide a complete view of the network under concern,
however how long the measurement is continued (for instance,
some links may be invisible from the sources used in
a {\tt traceroute}-based measurement).

Estimating such biases currently is a challenging area of
research in which some significant contributions have been
made \cite{lakhina02sampling,barford01imc,achlioptas05bias,guillaume2005metro,dallasta04statistical},
but most remains to be done.
The ultimate goal in this direction is to be able to accurately
evaluate the actual properties of a complex network from the
observation of a (biased) measurement.
In the absence of such results, researchers
have no choice but to rely on the assumption that the
properties they observe do not suffer from such a bias;
our method makes it possible to distinguish between cases
where this assumption is reasonable, and cases where it
must be discarded.

\medskip

Finally, two other observations obtained in this
contribution are worth pointing out.

\medskip

First,
it must be clear that the observed {\em qualitative} properties
are reliable: they do not depend on the sample size,
as long as it is not trivially small.
In particular, the average degree is small, the
density is close to $0$, the diameter and average distance are
small, the degree distributions are heterogeneous, and the clustering
coefficient and transitivity ratio are significantly larger than
the density (except for \ip, as explained in Section~\ref{sec_clustering}).
This is in full accordance with classical {\em qualitative} assumptions.

However, as discussed in Section~\ref{sec_context}, obtaining accurate
estimations of the {\em values} of the properties is crucial for modeling
and simulation: these values are used as key parameters in
these contexts and have significant impact on the obtained results.
Knowing the qualitative behavior of these properties therefore is
unsufficient, and our method constitutes a significant step
towards rigorously evaluating their actual values.

\medskip

Secondly, we gave strong evidence of the fact that the {\em evolution}
of many subtle statistics is well captured by the evolution of much
more basic statistics:
the average degree seems to control the general behavior of the
average distance and diameter, as well as the evolution of the
degree distribution,
and the transitivity ratio evolution seems to be governed by the ones
of the maximal degree
and density.
The more complex statistics are not totally controlled by simpler
ones, however, and investigating the difference between their behavior and what
can be expected would certainly yield enlightening insights.
In this spirit, we have shown that the ratio between the clustering coefficient
and the density seems significantly more stable than these two
statistics on their own.

These observations have to be deepened,
but they indicate that the set of relevant statistics for the study
of complex networks might be different from what is usually thought:
some statistics may be redundant, 
and other statistics may be more relevant than classical ones
(in particular, concerning their accurate evaluation).
This raises promising directions for further investigation, in both the
analysis and modeling areas.

\bigskip
\noindent
{\bf Acknowledgments.}
We thank all the colleagues who provided data to us, in particular
Paolo Boldi from WebGraph \cite{webgraphurl}, and the people at
MetroSec \cite{metrosecurl}, Skitter \cite{skitterurl}
and Lugdunum \cite{lugdunumurl}. No such work would be possible
without their help. We also thank Nicolas Larrieu for great help in
managing the data, and Fr\'ed\'eric Aidouni and Fabien Viger
for helpful comments and references.\\
This work was partly funded by the MetroSec (Metrology of the Internet for Security) \cite{metrosecurl},
and the AGRI (Analyse des Grands R\'eseaux d'Interactions)
projects.



\bibliographystyle{plain}
\bibliography{biblio}

\end{document}